\documentclass[prl,letterpaper,twocolumn,showpacs,superscriptaddress,amsmath,amsfonts,amssymb,preprintnumbers]{revtex4}
\pdfoutput=1 
\usepackage[T1]{fontenc}\usepackage[latin1]{inputenc}
\usepackage{dcolumn,graphicx,color,hvfloat,booktabs,microtype,afterpage}
\usepackage[loop,controls,buttonsize=2em]{animate}
\usepackage[charter]{mathdesign}
\renewcommand{\figurename}{Fig.}
\renewcommand{\tablename}{Table}
\makeatletter\renewcommand{\fnum@figure}[1]{\figurename~\thefigure~(color online).}\makeatother
\makeatletter\renewcommand{\fnum@table}[1]{\tablename~\thetable.}\makeatother

\newcount\hh \newcount\mm
\hh=\time \divide\hh by 60
\mm=\hh \multiply\mm by 60 \mm=-\mm
\advance\mm by \time
\def\now{\number\hh:\ifnum\mm<10{}0\fi\number\mm}

\usepackage[colorlinks,plainpages=false,linkcolor=blue,urlcolor=blue,citecolor=blue,pdfpagemode=UseNone,pdfstartview=FitBH]{hyperref}

\begin{document} \pagestyle{plain}

\title{Competing charge density waves and temperature-dependent nesting in $2H$-TaSe$_2$}

\author{Ph.~Leininger}\email[\vspace{-0.9ex}E-mail: \vspace{4pt}]{p.leininger@fkf.mpg.de}
\affiliation{Max Planck Institute for Solid State Research, Heisenbergstra{\ss}e~1, D-70569 Stuttgart, Germany.}

\author{D.~Chernyshov}
\affiliation{Swiss-Norwegian Beam Lines, European Synchrotron Radiation Facility (ESRF), BP 220, F-38043 Grenoble Cedex, France.}

\author{A.\,Bosak}
\affiliation{European Synchrotron Radiation Facility (ESRF), 6 rue Jules Horowitz, BP 220, F-38043 Grenoble Cedex, France.}

\author{H.~Berger}
\affiliation{Institut de Physique de la Mati\`{e}re Complexe, EPFL, 1015 Lausanne, Switzerland.}

\author{D.\,S.\,Inosov}\email[\vspace{-0.9ex}E-mail: \vspace{4pt}]{d.inosov@fkf.mpg.de}
\affiliation{Max Planck Institute for Solid State Research, Heisenbergstra{\ss}e~1, D-70569 Stuttgart, Germany.}



\begin{abstract}

\noindent Multiple charge density wave (CDW) phases in 2$H$-TaSe$_2$ are investigated by high-resolution synchrotron x-ray diffraction. In a narrow temperature range immediately above the commensurate CDW transition, we observe a multi-$\mathbf{q}$ superstructure with coexisting commensurate and incommensurate order parameters, clearly distinct from the fully incommensurate state at higher temperatures. This multi-$\mathbf{q}$ ordered phase, characterized by a temperature hysteresis, is found both during warming and cooling, in contrast to previous reports. In the normal state, the incommensurate superstructure reflection gives way to a broad diffuse peak that persists nearly up to room temperature. Its position provides a direct and accurate estimate of the Fermi surface nesting vector, which evolves non-monotonically and approaches the commensurate position as the temperature is increased. This behavior agrees with our recent observations of the temperature-dependent Fermi surface in the same compound [Phys.~Rev.~B \textbf{79}, 125112 (2009)].

\end{abstract}

\maketitle\enlargethispage{3pt}

\noindent The 2$H$ polymorph of tantalum diselenide (2$H$-TaSe$_2$) with a layered hexagonal crystal structure \cite{WilsonDiSalvo75} is among the most studied charge density wave (CDW) systems. It is a quasi-two-dimensional metal \cite{BorisenkoKordyuk08} known for its incommensurate (ICDW) and commensurate (CCDW) density-wave phases, which set in below $T_{\rm ICDW} \approx 120$\,K and $T_{\rm CCDW} \approx 90$\,K, respectively \cite{TemperatureNote}. In its ground state, this system is a superconductor with a rather low critical temperature, $T_{\rm c}=0.15$\,K \cite{GabovichVoitenko01}. Early neutron diffraction experiments \cite{MonctonAxe75} revealed incommensurate Bragg peaks close to the $(\frac{1}{\protect\raisebox{0.3pt}{\scriptsize 3}}\kern.5pt0\,0)$ and $(\frac{2}{\protect\raisebox{0.3pt}{\scriptsize 3}}\kern.5pt0\,0)$ reciprocal-space positions below the second-order ICDW transition. Only the longitudinal incommensurability, measured along the (100) direction, was investigated, and was shown to decrease monotonically with decreasing temperature, $T\!$, until it vanished abruptly at the first-order CCDW lock-in transition. However, x-ray diffraction measurements that followed shortly thereafter \cite{FlemingMoncton80} suggested the existence of another interjacent phase that appeared only upon warming at intermediate temperatures between $T_{\rm CCDW}$ and $T_{\rm C+IC} \approx 112$\,K, followed by the fully incommensurate phase at $T\!>\kern-1.5pt T_{\rm C+IC}$. It was ascribed to a ``striped-incommensurate'' triple-$\mathbf{q}$ ordering, characterized by one commensurate and two incommensurate propagation vectors.

In a recent angle-resolved photoelectron spectroscopy (ARPES) study \cite{KordyukBorisenko09, EvtushinskyKordyuk08, InosovEvtushinsky09, InosovZabolotnyy08}, effects of the static and fluctuating CDW on the temperature dependence of the 2$H$-TaSe$_2$ Fermi surface (FS) have been investigated. In particular, the normal-state FS has been mapped with unprecedented accuracy owing to its highly two-dimensional character. The most striking output of these experiments was the observation of a non-monotonic $T\!$-dependence of the pseudogap \cite{BorisenkoKordyuk08, KordyukBorisenko09, EvtushinskyKordyuk08}, accompanied by a subtle, similarly non-monotonic variation of the FS \cite{BorisenkoKordyuk08, InosovEvtushinsky09}. On the one hand, the persistence of these effects into the normal state, well above the ordering temperature, suggested that CDW fluctuations must be involved. On the other hand, the non-monotonic $T\!$-dependence could not be explained by a simple picture of critical fluctuations of the ICDW order that would rapidly decay above $T_{\rm ICDW}$, leading to a monotonic closing of the pseudogap. Therefore, a more intricate feedback effect of these fluctuations on the nesting properties of the FS could be suspected. This served as our motivation to take a closer look at the competition of commensurate and incommensurate CDW order parameters and to study the $T$-dependence of the nesting vector in the normal state.

Being a highly accurate reciprocal-space probe of the band structure, ARPES can still measure the FS nesting properties only indirectly, e.g.~via calculations of the particle-hole correlation functions based on fitting a tight-binding model to the measured spectra \cite{InosovZabolotnyy08, InosovBorisenko07}. The accuracy of such analysis is insufficient to quantify the tiny incommensurability of the nesting vector, which in the case of 2$H$-TaSe$_2$ remains below 1\% of the unfolded Brillouin zone (BZ) size at any temperature \cite{FlemingMoncton80}.

In the present letter, we address this challenge using high-resolution synchrotron x-ray diffraction. Not only does it provide the necessary momentum resolution for measurements of the incommensurability, but also serves as a more direct probe of the nesting vector due to the appearance of a thermal diffuse-scattering signal in the normal state \cite{WilsonDiSalvo75, ThermalDiffuseScattering}. Here we systematically investigate its temperature dependence using x-rays with a wavelength of $\lambda=0.7$\,\AA\ ($\hbar\omega=17.7$\,keV), characterized by a bandwidth of $\Delta\lambda/\lambda\approx2\times\!10^{-4}$. A platelike 20-$\mu$m-thick single crystal of 2$H$-TaSe$_2$ was cleaved mechanically from a bigger crystal and mounted on a rotation stage with its $\mathbf{c}$-axis parallel to the incident beam. The sample was then cooled down to 85\,K with a \textit{Cryostream 700+} nitrogen cooler (\textit{Oxford Cryosystems}). The data were measured in transmission geometry using a multi-purpose KUMA KM6-CH diffractometer and collected by successive frames with an \textit{Onyx} CCD area detector. The orientation matrix was determined with the CRYSALIS software package (\textit{Oxford Diffraction, Ltd.}).

\begin{figure*}[t]\vspace{-1em}
\includegraphics[width=\textwidth]{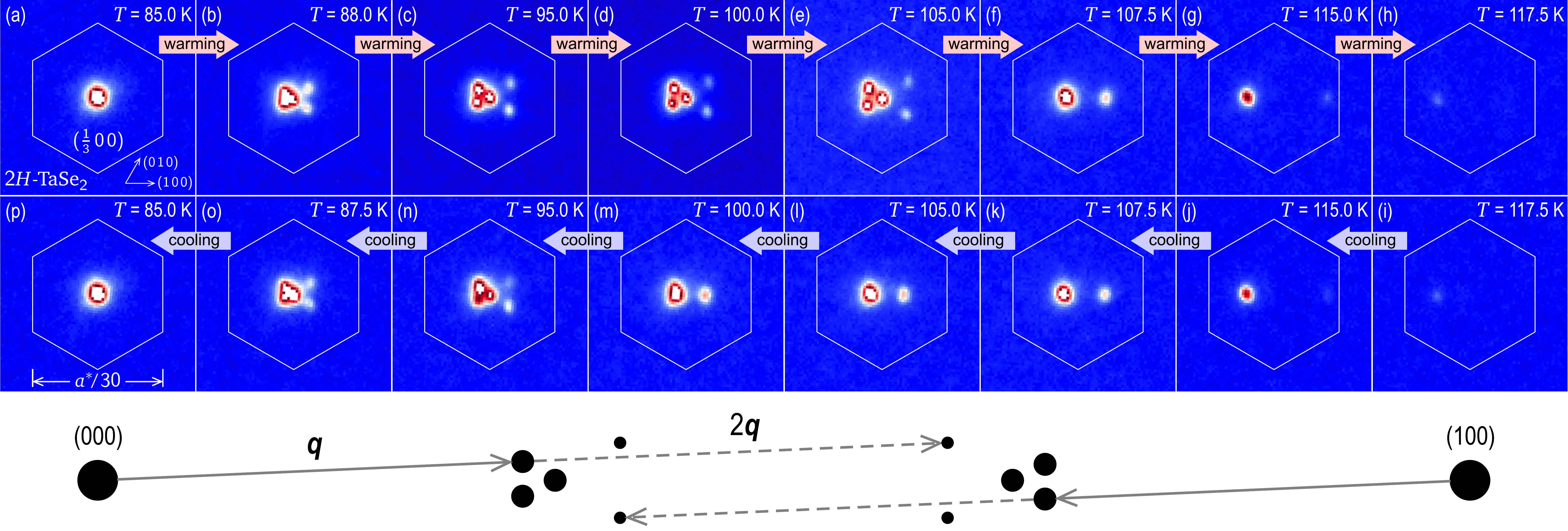}\vspace{-0.5em}
\caption{Temperature evolution of the $(\frac{1}{\protect\raisebox{0.3pt}{\scriptsize 3}}\kern.5pt0\,0)$ superstructure reflection across $T_{\rm CCDW}$ and $T_{\rm C+IC}$ during warming (top) and cooling (bottom) \cite{Supplement}. The color represents diffracted intensity. Note the hysteretic behavior between 95 and 105\,K. The momentum scale is given by the white hexagons, which are scaled to 1/10 of the folded BZ size. The schematic below the figure explains the origin of second-order incommensurate reflections in the multi-$\mathbf{q}$ CDW phase. Here the incommensurability is largely exaggerated for clarity. For an animated version of this figure with additional temperature frames, see supplementary material \cite{Supplement}.\vspace{-0.8em}}
\label{Fig:Bragg}
\end{figure*}

\begin{figure}[b]\vspace{-0.1em}
\includegraphics[width=\columnwidth]{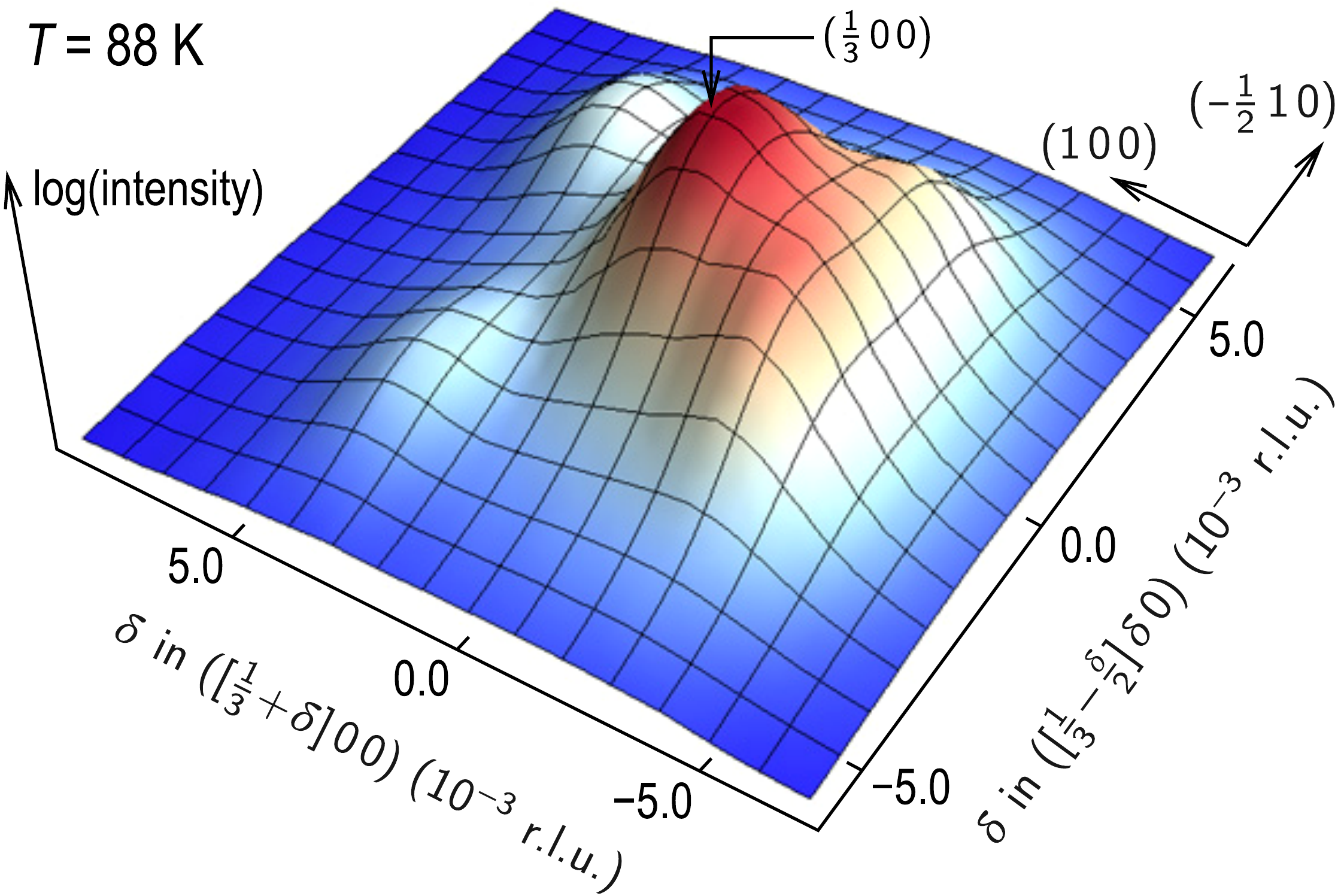}\vspace{-0.5em}
\caption{Distribution of the superstructure Bragg intensity, measured upon warming around $(\frac{1}{\protect\raisebox{0.3pt}{\scriptsize 3}}\kern.5pt0\,0)$ at $T=88$\,K and integrated over several frames along the sample rocking angle. Note the logarithmic intensity scale.\vspace{-1.4em}}
\label{Fig:SurfacePlot}
\end{figure}

At first, we consider the temperature dependence of the superstructure reflections across the ordered phases, measured in the vicinity of the $(\frac{1}{\protect\raisebox{0.3pt}{\scriptsize 3}}\kern.5pt0\,0)$ reciprocal-lattice vector (see Fig.\,\ref{Fig:Bragg} or Fig.\,S1\,(a) in the online supplementary material \cite{Supplement}). In the low-temperature CCDW phase, a commensurate superstructure reflection is observed at $(\frac{1}{\protect\raisebox{0.3pt}{\scriptsize 3}}\kern.5pt0\,0)$, as expected. Above 87\,K, it splits into five closely spaced peaks (Fig.\,\ref{Fig:SurfacePlot}), one of which remains at the commensurate position, whereas the others represent first- and second-order incommensurate superstructure reflections that appear at $(\frac{1}{\protect\raisebox{0.3pt}{\scriptsize 3}}\kern-.5pt-\kern-.5pt\delta,\,\pm\frac{1}{\sqrt{3}}\delta,~0)$ and $(\frac{1}{\protect\raisebox{0.3pt}{\scriptsize 3}}\kern-.5pt+\kern-.5pt2\delta,\,\pm\frac{2}{\sqrt{3}}\delta,~0)$ reciprocal-space vectors, respectively.

The individual frames in Fig.\,\ref{Fig:Bragg} were measured at a fixed sample position and therefore can not accurately represent the full peak intensities. Therefore, we have integrated the diffracted intensity within several such frames, measured at different rocking angles, to get more reliable intensity distributions. The surface plot in Fig.\,\ref{Fig:SurfacePlot} shows one such integrated pattern at $T=88$\,K, immediately above the CCDW transition, where the development of four incommensurate satellites can be already clearly seen, though not yet fully resolved. Their positions agree with those evaluated from Fig.\,\ref{Fig:Bragg}\,(b). The origin of these multiple peaks is explained in Ref.\,\onlinecite{FlemingMoncton80} as a result of orthorhombic twinning of a triple-$\mathbf{q}$ superstructure with one commensurate and two incommensurate propagation vectors (C+IC), oriented at nearly 120$^\circ$ to each other. The relative intensities of the peaks are therefore determined by the twin-domain population. The appearance of second-order reflections is explained in the schematic at the bottom of Fig.\,\ref{Fig:Bragg}.

\begin{figure*}[t]\vspace{-1ex}
\includegraphics[width=\textwidth]{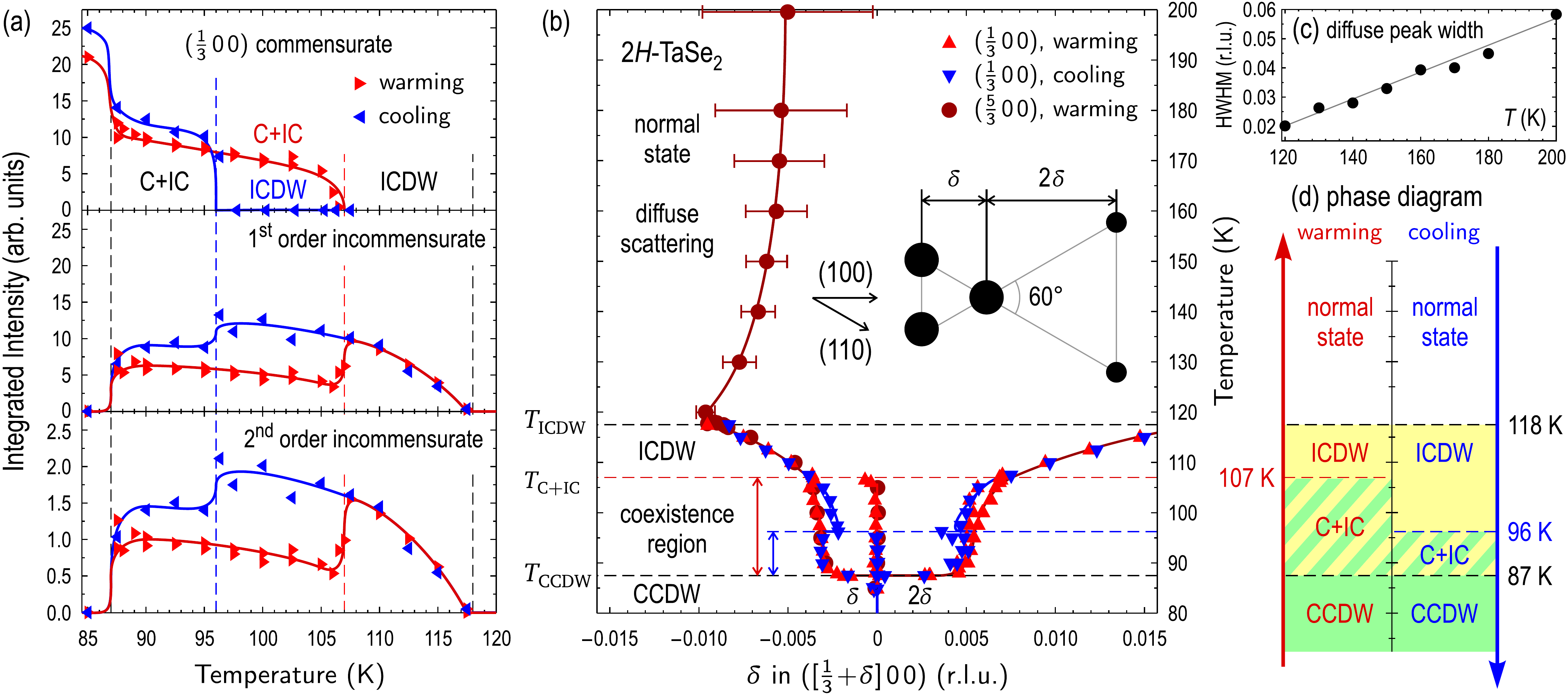}\vspace{-0.5em}
\caption{\textbf{(a)}~Intensities of the commensurate (top) and incommensurate (middle and bottom) superstructure peaks as functions of temperature, measured during warming and cooling. In the C+IC phase, the sum of intensities for both incommensurate peaks is shown. Solid lines are guides to the eyes. \textbf{(b)}~$T\!$-dependence of the longitudinal incommensurability, $\delta$. The inset clarifies the definition of $\delta$ in the multi-$\mathbf{q}$ CDW phase. \textbf{(c)}~Half width at half maximum of the diffuse peak in the normal state as a function of $T$. \textbf{(d)}~Summary of transition temperatures between the normal state, ICDW, C+IC, and CCDW phases upon cooling and warming.\vspace{-1.0em}}
\label{Fig:Incommensurability}
\end{figure*}

To extract the peak positions and intensities (Fig.\,\ref{Fig:Incommensurability}), we have fitted the data with a sum of two-dimensional Gaussian peaks. The intensity ratio between the first- and second-order diffraction peaks was roughly evaluated to 7 and was kept constant for all the fits between T$_{\rm ICDW}$ and T$_{\rm CCDW}$, whereas the peak widths, positions and intensities were fitted as free parameters. Within the accuracy of our measurement, the incommensurate Bragg reflections always appear symmetrically with respect to the $(110)$ direction of the folded BZ, so that the first-order reflections in the C+IC phase form an equilateral triangle with its vertex at $(\frac{1}{\protect\raisebox{0.3pt}{\scriptsize 3}}\kern.5pt0\,0)$. Upon warming above $T_{\rm C+IC\,\uparrow}\approx107$\,K, the commensurate peak disappears, and at the same time all four incommensurate peaks collapse onto the $(100)$ axis, losing their transverse incommensurability but preserving its longitudinal component, $\delta$, without any discontinuous changes (Figs.\,\ref{Fig:Bragg} and \ref{Fig:Incommensurability}). This signifies the first-order transition to the fully incommensurate state that persists up to $T_{\rm ICDW}\approx117$\,K. Above this temperature, broad diffuse-scattering peaks can be seen \cite{Supplement}.

In contrast to Ref.\,\onlinecite{FlemingMoncton80}, we observe the C+IC phase also upon cooling below $T_{\rm C+IC\,\downarrow}\approx96$\,K. The phase transition at $T_{\rm C+IC\,\downarrow}$ exhibits a pronounced temperature hysteresis, in contrast to the lock-in transition at $T_{\rm CCDW}$ that does not depend on the direction of the temperature change. This is best seen in the $T\!$-dependent intensity of the superstructure reflections shown in Fig.\,\ref{Fig:Incommensurability}\,(a). The corresponding $T\!$-dependence of the longitudinal incommensurability and the temperature phase diagram are summarized in Fig. \ref{Fig:Incommensurability}\,(b) and (d), respectively. Note that the C+IC phase is clearly distinct from a simple superposition of the CCDW and ICDW phases (which could occur as a result of phase separation, for example). Indeed, the incommensuration in the C+IC phase occurs along the (110) direction, whereas in the ICDW phase only the longitudinal component is preserved. This implies an interplay between three competing CDW vectors that possibly originate from the multiple local minima in the electronic nesting function \cite{InosovZabolotnyy08, InosovEvtushinsky09}. On the other hand, the fact that the incommensurability experiences a discontinuous jump at $T_{\rm C+IC}$ only in the transverse direction, whereas its longitudinal component changes continuously across the whole phase diagram, implies a close relationship between these order parameters, with a very low energy barrier separating the two incommensurate ground states.

\begin{figure}[b]\vspace{-0.1em}
\includegraphics[width=\columnwidth]{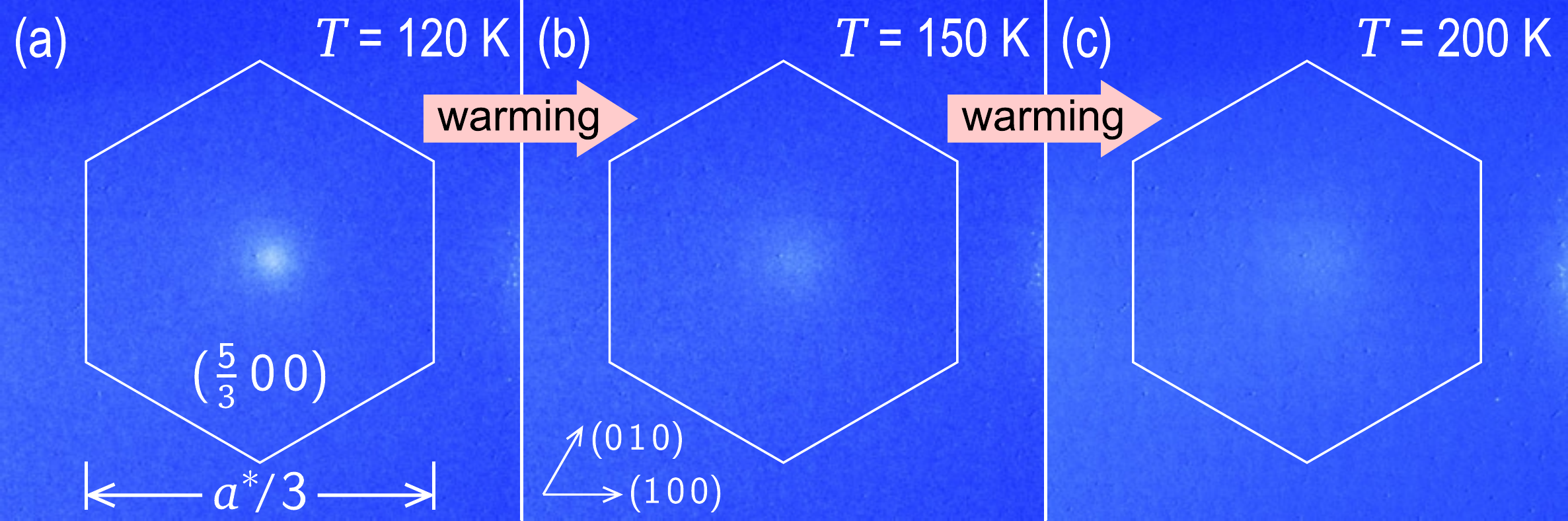}\vspace{-0.5em}
\caption{Temperature dependence of the diffuse-scattering intensity measured near the $(\frac{5}{\protect\raisebox{0.3pt}{\scriptsize 3}}\kern.5pt0\,0)$ reciprocal-space vector. The white hexagons circumscribe the folded BZ boundary (i.e. are 10 times larger than in Fig.\,\ref{Fig:Bragg}). For an animated version of this figure with additional temperature frames, see Ref.\,\onlinecite{Supplement}.\vspace{-1.4em}}
\label{Fig:Diffuse}
\end{figure}

In the normal state above $T_{\rm ICDW}$, the first-order superstructure reflection gives way to a broad diffuse peak of nearly isotropic circular shape (Fig.\,\ref{Fig:Diffuse}), which gets weaker and broadens linearly with increasing temperature, as shown in Fig.\,\ref{Fig:Incommensurability}\,(c). This peak can originate either from the inelastic scattering on the soft phonon mode \cite{MonctonAxe75, ThermalDiffuseScattering} or from the quasielastic scattering on the charge-density fluctuations, which are both peaked at the Fermi-surface nesting vector. We have measured the positions of this diffuse peak in a higher BZ near the $(\frac{5}{\protect\raisebox{0.3pt}{\scriptsize 3}}\kern.5pt0\,0)$ wavevector to minimize the background intensity from the direct beam. These peak positions, extracted by fitting the diffraction patterns to a two-dimensional Gaussian profile with a linear background, mostly related to residual scattering from the incident beam, are shown in Fig.\,\ref{Fig:Incommensurability}\,(b) as circles with error bars. At $T_{\rm ICDW}$, the center of the diffuse peak coincides with the incommensurate superstructure reflection of the ICDW phase. At higher temperatures, however, the diffuse peak gradually shifts towards the commensurate wavevector. Hence, a pronounced cusp in the $T$-dependence of $\delta$ is formed around $T_{\rm ICDW}$. A similar behavior has been previously reported at the ICDW transition in SmNiC$_2$ \cite{ShimomuraHayashi09}. Although such a non-monotonic behavior is fully consistent with the fluctuation-driven temperature variations of the FS observed in 2$H$-TaSe$_2$ earlier \cite{BorisenkoKordyuk08, InosovEvtushinsky09}, it is not captured by the conventional theory of density-wave phase transitions \cite{ChanHeine73} that considers the normal-state FS temperature-independent and fully determined by the static lattice potential.

We conclude that a proper theoretical description of density-wave systems must take the dynamic fluctuation effects into account. This necessity is dictated by the ever-increasing accuracy of the experimental probes that can no longer be reconciled with a conventional (static) description. As we have shown, these effects can lead to experimentally observable deformations of the FS, which, in their turn, exhibit a subtle feedback effect on the position of the nesting vector that determines the ordering instability. A self-consistent description is therefore required to match the FS nesting vector with the CDW ordering wavevector within a comprehensive theoretical model.

\textit{Acknowledgements.} The x-ray diffraction measurements reported here were performed at the Swiss-Norwegian beamline BM1A of the European Synchrotron Radiation Facility (ESRF). We are grateful to S.\,V.~Borisenko, D.\,V.~Efremov, D.\,V.~Evtushinsky, A.\,Gabovich, V.~Hinkov, T.~Keller, A.\,A.\,Kordyuk, N.\,Munnikes, M.\,Le~Tacon, I.\,Zegkinoglou and B.\,Keimer for discussions and helpful suggestions.\clearpage

\makeatletter\renewcommand{\fnum@figure}[1]{\figurename~S\thefigure.}\makeatother
\setcounter{figure}{0}
\pagestyle{plain}
\makeatletter
\renewcommand{\@oddfoot}{\hfill\bf\scriptsize\textsf{~}}
\renewcommand{\@evenfoot}{\bf\scriptsize\textsf{~}\hfill}
\makeatother
\onecolumngrid
\begin{center}{\large Supplementary material for the letter entitled\bigskip\\
\textbf{``Competing charge density waves and temperature-dependent nesting in $2H$-TaSe$_2$''}\bigskip\\
by Ph.~Leininger, D.~Chernyshov, A.\,Bosak, H.~Berger, and D.\,S.\,Inosov}\end{center}\bigskip

\begin{figure}[h]\label{supplement}
\raisebox{0.43\textwidth}{\Large (a)~}\animategraphics[timeline="bragg.timeline", width=0.45\textwidth]{2}{bragg_}{1}{35}\hfill
\raisebox{0.43\textwidth}{\Large (b)~}\animategraphics[timeline="diffuse.timeline", width=0.45\textwidth]{2}{diffuse_}{1}{20}
\caption{Animated temperature dependence of diffracted intensity in 2$H$-TaSe$_2$. Use controls below the figure to start the animations, change the frame rate or navigate through individual frames. (a)~Evolution of superstructure Bragg reflections in the vicinity of $(\frac{1}{\protect\raisebox{0.3pt}{\scriptsize 3}}\kern.5pt0\,0)$ during warming and cooling. The momentum scale is given by the white hexagon, which is scaled to 1/10 of the folded BZ size. (b)~Appearance of the diffuse scattering near the $(\frac{5}{\protect\raisebox{0.3pt}{\scriptsize 3}}\kern.5pt0\,0)$ superstructure reflection in the normal state. Here the white hexagon circumscribes the folded BZ boundary, i.e. is 10 times larger than in panel (a).}
\label{Animation}
\end{figure}


\begin{thebibliography}{10}
\bibitem{TemperatureNote} The transition temperatures observed in the present study are consistent with those determined in the early diffraction experiments (Ref.\,\onlinecite{WilsonDiSalvo75}), but are somewhat lower than those observed by neutron scattering (Ref.\,\onlinecite{MonctonAxe75}).
\end{thebibliography}
\end{document}